# Large System Analysis of Power Normalization Techniques in Massive MIMO

Meysam Sadeghi, *Student Member, IEEE*, Luca Sanguinetti, *Senior Member, IEEE*, Romain Couillet, *Senior Member, IEEE*, and Chau Yuen, *Senior Member, IEEE*

*Abstract*—Linear precoding has been widely studied in the context of Massive multiple-input-multiple-output (MIMO) together with two common power normalization techniques, namely, matrix normalization (MN) and vector normalization (VN). Despite this, their effect on the performance of Massive MIMO systems has not been thoroughly studied yet. The aim of this paper is to fulfill this gap by using large system analysis. Considering a system model that accounts for channel estimation, pilot contamination, arbitrary pathloss, and per-user channel correlation, we compute tight approximations for the signal-to-interference-plus-noise ratio and the rate of each user equipment in the system while employing maximum ratio transmission (MRT), zero forcing (ZF), and regularized ZF precoding under both MN and VN techniques. Such approximations are used to analytically reveal how the choice of power normalization affects the performance of MRT and ZF under uncorrelated fading channels. It turns out that ZF with VN resembles a sum rate maximizer while it provides a notion of fairness under MN. Numerical results are used to validate the accuracy of the asymptotic analysis and to show that in Massive MIMO, non-coherent interference and noise, rather than pilot contamination, are often the major limiting factors of the considered precoding schemes.

## I. INTRODUCTION

Massive multiple-input multiple-output (MIMO) is a multiuser MIMO system that employs a large number of antennas at the base stations (BSs) to serve a relatively smaller number of user equipments (UEs) [1]–[4]. This large number of antennas enables each BS to focus the radiated energy into a specific location in space or to intercept the power of transmitted electromagnetic waves more efficiently. Therefore, Massive MIMO has higher spectral efficiency and energy efficiency compared to classical multiuser MIMO systems [3], [5]–[7]. Due to the quasi-orthogonal nature of channels in Massive MIMO, linear precoding and detection schemes perform close-to-optimal [5], [6], [8]. If the channel reciprocity is exploited, the overhead of the channel state information (CSI) acquisition is independent of the number of BS antennas [9]. Moreover, it has been recently shown that the capacity of Massive MIMO increases without bound as the number of antennas increases, even under pilot contamination [10], [11]. These remarkable features candidate Massive MIMO as one of the most promising technologies for next generation of cellular networks [8], [12], [13].

Linear precoding has a central role in Massive MIMO and has been extensively studied in the past few years [5], [14]–[22]. The spectral efficiency and energy efficiency of maximum ratio transmission (MRT) and zero forcing (ZF) precoding in single-cell Massive MIMO systems are investigated in [14]. In [15], a multicell linear precoding is proposed to mitigate the effect of pilot contamination. Multicell processing is also considered in [10], [11], [21]. The performance of MRT, ZF, and regularized ZF (RZF) precoding in single-cell large-scale MIMO systems is studied in [16], considering a per-user channel correlation model. A seminal treatment of MRT and RZF precoding schemes in multicell Massive MIMO systems is presented in [5], followed by [17] where downlink training and linear pilot contamination precoding are also considered. In [18], closed-form approximations for the achievable downlink rates of MRT and ZF precoding schemes are presented for multicell Massive MIMO systems. A linear truncated polynomial expansion based precoding is proposed in [19], which reduces the complexity of RZF precoding. The effect of phase noise on the signal-to-interference-plus-noise (SINR) of MRT, ZF, and RZF precoding schemes is studied in [20].

In order to utilize linear precoding, the power should be adjusted to meet the power constraint at the BS. This can be done either by optimized power allocation among the downlink data streams [21], [23]–[25], or simply by uniform power allocation among downlink data streams jointly with precoder power normalization [5], [14], [17]–[19]. Although the latter approach may provide a weaker performance compared to the former, it is the most used in the Massive MIMO literature [5], [9], [14], [17]–[19]. The reason for this is that power allocation presents the following major issues: ($i$) finding a global solution is a challenging task [24], [26], [27]; ($ii$) a certain level of coordination or cooperation among cells is required; and ($iii$) it should be performed very frequently, even for static users, as scheduling may change rapidly in practice.

The two commonly used power normalization techniques

M. Sadeghi (meysam@mymail.sutd.edu.sg) and C. Yuen (yuenchau@sutd.edu.sg) are with Singapore University of Technology and Design (SUTD), Singapore. L. Sanguinetti (luca.sanguinetti@unipi.it) is with the University of Pisa, Dipartimento di Ingegneria dell'Informazione, Italy and also with the Large Systems and Networks Group (LANEAS), CentraleSupélec, Université Paris-Saclay, 3 rue Joliot-Curie, 91192 Gif-sur-Yvette, France. R. Couillet (romain.couillet@centralesupelec.fr) is with the Signals and Statistics Group, CentraleSupélec, Université Paris-Saclay, 4103 Paris, France.

This work was supported by A*Star SERC project number 142-02-00043. L. Sanguinetti and R. Couillet have been supported by the ERC Starting Grant 305123 MORE.

in Massive MIMO are matrix normalization (MN) and vector normalization (VN) [26], [27]. In MN, the precoding matrix of each BS is adjusted by multiplying it with a scalar such that the power constraint at the BS is met [5], [9], [14], [17], [19]. On the other hand, with VN the precoding matrix is normalized such that equal amount of power is allocated to each UE while satisfying the power constraint [18], [26], [27]. Note that these two methods yield the same performance with optimal power allocation, but not with practical suboptimal power allocation [27], [28].

Although linear precoding has been largely studied in Massive MIMO, a detailed treatment of the impact of power normalization does not exist in the literature. The first attempt in this direction was carried out in [26] and extended in [27] wherein the authors study the impact of MN and VN on MRT and ZF precoding schemes. However, both [26], [27] do not grasp the essence of a practical Massive MIMO system since: ($i$) a single-cell network composed of three radio units is considered; ($ii$) perfect CSI is assumed and thus CSI acquisition or pilot contamination are not accounted for; and ($iii$) large-scale attenuation is neglected, though it has a fundamental impact on power normalization, as detailed later.

The goal of this paper is to study the effect of MN and VN on the performance of MRT, ZF, and RZF in Massive MIMO, in the simple and practical case of uniform power allocation. Particularly, the following contributions are provided.

- We extend the analysis in [26], [27] to a multicell Massive MIMO system, which accounts for channel estimation, pilot contamination, an arbitrary pathloss model, and per-user channel correlation. Asymptotically tight approximations of the signal-to-interference-plus-noise ratio (SINR) and rate of each UE are provided and validated by numerical results for MRT, ZF, and RZF with VN and MN.
- Explicit asymptotic approximations for the SINR and rate of each UE are given for a Rayleigh fading channel model. These results are used: ($i$) to elaborate on how the two different normalization techniques affect the signal, noise, and interference powers as well as the pilot contamination experienced by each UE in the system; ($ii$) to prove that large-scale fading has a fundamental role on the performance provided by the two normalization techniques while both perform the same if neglected; ($iii$) to show that ZF conveys a notion of sum rate maximization with VN and of fairness with MN.
- The asymptotic approximations of SINRs are used together with numerical results to study the main limiting factors of the investigated schemes. Particularly, we reveal that in Massive MIMO, non-coherent interference and noise, rather than pilot contamination, are often the major limiting factors for all schemes.

The remainder of this paper is organized as follows. Section II introduces the network model, the channel estimation scheme, the precoding and power normalization methods, as well as the downlink achievable rates. The large system analysis is provided in Section III. The effect of power normalization techniques is elaborated in Section IV for uncorrelated fading channels. The provided asymptotic approximations are verified by means of numerical results in Section V. Some conclusions are drawn in Section VI.

*Notations:* The following notation is used throughout the paper. Scalars are denoted by lower case letters whereas boldface lower (upper) case letters are used for vectors (matrices). We denote by $\mathbf{I}_N$ the identity matrix of size $N$ and represent the element on the $i$th row and $k$th column of $\mathbf{A}$ as $[\mathbf{A}]_{i,k}$. The symbol $\mathcal{CN}(.,.)$ denotes the circularly symmetric complex Gaussian distribution. The trace, transpose, conjugate transpose, real part, and expectation operators are denoted by $\mathsf{tr}(\cdot)$, $(\cdot)^T$, $(\cdot)^H$, $\mathsf{Re}(\cdot)$, and $\mathbb{E}[\cdot]$, respectively. The notation $\xrightarrow[N\to\infty]{}$ represent almost sure convergence.

## II. COMMUNICATION SCHEME

Next, we introduce the system model, the channel estimation method, the precoding and power normalization techniques, and compute the downlink achievable rates.

### A. System Model

We consider the downlink of a Massive MIMO system composed of $L$ cells, where the set of all cells is denoted by $\mathcal{L}$. The BS of each cell has $N$ antennas and serves $K$ single-antenna UEs in the same time-frequency resource. The set of UEs belonging to cell $l$ is denoted by $\mathcal{K}_l$. We assume transmissions over flat-fading channels. We employ a double index notation to refer to each UE as e.g., "user $k$ in cell $l$". Under this convention, let $\mathbf{h}_{jlk} \in \mathbb{C}^N$ be the channel between BS $j$ and UE $k$ in cell $l$ within a block and assume that

$$\mathbf{h}_{jlk} = \boldsymbol{\Theta}_{jlk}^{1/2} \mathbf{z}_{jlk} \quad (1)$$

where $\mathbf{z}_{jlk} \sim \mathcal{CN}(\mathbf{0}, \mathbf{I}_N)$ and $\boldsymbol{\Theta}_{jlk} \in \mathbb{C}^{N \times N}$ accounts for the corresponding channel correlation matrix. Note that (1) enables us to assign a unique correlation matrix between each user-BS pair and it includes many channel models in the literature as special cases [16].

### B. Channel Estimation

We assume that BSs and UEs are perfectly synchronized and operate according to a time-division duplex (TDD) protocol. Then, the channels can be found by an uplink training phase and used in the downlink by exploiting channel reciprocity. Using orthogonal pilots in each cell while reusing them in all other cells, after correlating the received training signal with the pilot sequence of UE $k$, the observed channel of user $k$ in cell $j$ is

$$\mathbf{y}_{jk}^{\mathrm{tr}} = \mathbf{h}_{jjk} + \sum_{l=1, l \neq j}^{L} \mathbf{h}_{jlk} + \frac{1}{\sqrt{\rho_{\mathrm{tr}}}} \mathbf{n}_{jk} \quad (2)$$

where $\mathbf{n}_{jk} \sim \mathcal{CN}(\mathbf{0}, \sigma^2 \mathbf{I}_N)$ with $\sigma^2$ being the noise spectral density and $\rho_{\mathrm{tr}}$ is proportional to the training SNR. Applying the MMSE estimation, the estimated channel can be computed as follows [5]

$$\hat{\mathbf{h}}_{jjk} = \boldsymbol{\Theta}_{jjk} \mathbf{Q}_{jk} \mathbf{y}_{jk}^{\mathrm{tr}} \quad (3)$$

where $\hat{\mathbf{h}}_{jjk} \sim \mathcal{CN}(\mathbf{0}, \boldsymbol{\Phi}_{jjk})$. Also, $\mathbf{Q}_{jk}$ and $\boldsymbol{\Phi}_{jlk}$ are given by

$$\mathbf{Q}_{jk} = \left(\sum_{l=1}^{L} \boldsymbol{\Theta}_{jlk} + \frac{\sigma^2}{\rho_{\text{tr}}}\mathbf{I}_N\right)^{-1} \quad \forall j, k \quad (4)$$

$$\boldsymbol{\Phi}_{jlk} = \boldsymbol{\Theta}_{jjk} \mathbf{Q}_{jk} \boldsymbol{\Theta}_{jlk} \quad \forall j, l, k. \quad (5)$$

Note that due to the orthogonality principle of MMSE, the estimation error $\widetilde{\mathbf{h}}_{jjk} = \mathbf{h}_{jjk} - \hat{\mathbf{h}}_{jjk}$ is independent of $\hat{\mathbf{h}}_{jjk}$ and such that $\widetilde{\mathbf{h}}_{jjk} \sim \mathcal{CN}(\mathbf{0}, \boldsymbol{\Theta}_{jjk} - \boldsymbol{\Phi}_{jjk})$. For notational simplicity, we denote $\hat{\mathbf{H}}_{jj} = [\hat{\mathbf{h}}_{jj1}, \ldots, \hat{\mathbf{h}}_{jjK}]$ as the matrix collecting the estimated channels of cell $j$.

### C. Precoding and Power Normalization Techniques

As mentioned earlier, we consider MRT, ZF, and RZF with VN and MN [26], [27]. Denoting by $\mathbf{G}_j = [\mathbf{g}_{j1}, \ldots, \mathbf{g}_{jK}] \in \mathbb{C}^{N \times K}$ the precoding matrix of BS $j$, where $\mathbf{g}_{jk} \in \mathbb{C}^N$ is the precoding vector of UE $k$ in cell $j$, we have

$$\mathbf{G}_j = \mathbf{F}_j \mathbf{D}_j^{1/2} \quad (6)$$

where $\mathbf{F}_j = [\mathbf{f}_{j1}, \ldots, \mathbf{f}_{jK}] \in \mathbb{C}^{N \times K}$ determines the precoding scheme and $\mathbf{D}_j \in \mathbb{C}^{K \times K}$ characterizes the power allocation strategy. Therefore, $\mathbf{F}_j$ takes one of the following forms:

$$\mathbf{F}_j = \begin{cases} \hat{\mathbf{H}}_{jj} & \text{MRT} \quad (7) \\ \hat{\mathbf{H}}_{jj}\left(\hat{\mathbf{H}}_{jj}^H \hat{\mathbf{H}}_{jj}\right)^{-1} & \text{ZF} \quad (8) \\ \left(\hat{\mathbf{H}}_{jj}\hat{\mathbf{H}}_{jj}^H + \mathbf{Z}_j + N\alpha_j \mathbf{I}_N\right)^{-1}\hat{\mathbf{H}}_{jj} & \text{RZF} \quad (9) \end{cases}$$

where $\alpha_j > 0$ is the regularization parameter and $\mathbf{Z}_j \in \mathbb{C}^{N \times N}$ is an arbitrary Hermitian nonnegative definite matrix that can be used to leverage the system performance [5].

As mentioned in the introduction, finding the optimal values for the elements of $\mathbf{D}_j$ is challenging in practice [24]. This is why VN or MN are usually employed [27]. In this case, $\mathbf{D}_j$ is diagonal with entries chosen so as to satisfy the following average power constraint $\mathbb{E}[\text{tr}\mathbf{G}_j\mathbf{G}_j^H] = K \; \forall j$. If VN is used, then the $k$th diagonal element of $\mathbf{D}_j$ is computed as

$$[\mathbf{D}_j]_{k,k} = d_{jk} = \frac{1}{\mathbb{E}[\mathbf{f}_{jk}^H \mathbf{f}_{jk}]}. \quad (10)$$

On the other hand, if MN is employed, then $\mathbf{D}_j = \eta_j \mathbf{I}_K$ with

$$\eta_j = \frac{K}{\mathbb{E}[\text{tr}\mathbf{F}_j\mathbf{F}_j^H]}. \quad (11)$$

### D. Downlink Achievable Rate

The received signal of user $k$ in cell $j$ can be written as

$$y_{jk} = \mathbf{h}_{jjk}^H \mathbf{g}_{jk} s_{jk} + \sum_{i=1, i \neq k}^{K} \mathbf{h}_{jjk}^H \mathbf{g}_{ji} s_{ji} + \sum_{l=1, l \neq j}^{L} \sum_{i=1}^{K} \mathbf{h}_{ljk}^H \mathbf{g}_{li} s_{li} + n_{jk} \quad (12)$$

with $s_{li} \in \mathbb{C}$ being the signal intended to UE $i$ in cell $l$, assumed independent across $(l, i)$ pairs, of zero mean and unit variance, and $n_{jk} \sim \mathcal{CN}(0, \sigma^2/\rho_{\text{dl}})$ where $\rho_{\text{dl}}$ is proportional to the downlink signal power.

As in [1], [5], [6], [15] (among many others), we assume that there are no downlink pilots such that the UEs do not have knowledge of the current channels but can only learn the average channel gain $\mathbb{E}\{\mathbf{h}_{jjk}^H \mathbf{g}_{jk}\}$ and the total interference power. Note this is the common approach in Massive MIMO due to the channel hardening [29]. Using the same technique as in [30], an ergodic achievable information rate for UE $k$ in cell $j$ is obtained as $r_{jk} = \log_2(1 + \gamma_{jk})$ where $\gamma_{jk}$ is given by

$$\gamma_{jk} = \frac{|\mathbb{E}[\mathbf{h}_{jjk}^H \mathbf{g}_{jk}]|^2}{\frac{\sigma^2}{\rho_{\text{dl}}} + \sum_{l=1}^{L}\sum_{i=1}^{K} \mathbb{E}[|\mathbf{h}_{ljk}^H \mathbf{g}_{li}|^2] - |\mathbb{E}[\mathbf{h}_{jjk}^H \mathbf{g}_{jk}]|^2} \quad (13)$$

where the expectation is taken with respect to the channel realizations. The above result holds true for any precoding scheme and is obtained by treating the interference (from the same and other cells) and channel uncertainty as worst-case Gaussian noise. By using VN and MN, i.e. (10) and (11), the SINR takes respectively the form in (16) and (17), given on the top of next page.

As for all precoding schemes, $\gamma_{jk}^{MN}$ and $\gamma_{jk}^{VN}$ depend on the statistical distribution of $\{\mathbf{h}_{jlk}\}$ and $\{\hat{\mathbf{h}}_{jlk}\}$. This makes hard to compute both in closed-form. To overcome this issue, a large system analysis is provided next to find tight asymptotic approximations (hereafter called deterministic equivalents) for $\gamma_{jk}^{MN}$ and $\gamma_{jk}^{VN}$ and their associated achievable rates.

## III. LARGE SYSTEM ANALYSIS

We consider a regime in which $N$ and $K$ grow large with a non-trivial ratio $N/K$, where $1 < \liminf N/K \leq \limsup N/K < \infty$. We will represent it as $N \to \infty$. Under this assumption, we provide asymptotic approximations, also called deterministic equivalents (DEs), for $\gamma_{jk}$ with MRT, ZF, and RZF and either MN or VN. The DE is represented by $\overline{\gamma}_{jk}$, and it is such that $\gamma_{jk} - \overline{\gamma}_{jk} \xrightarrow[N \to \infty]{} 0$. By applying the continuous mapping theorem [31], the almost sure convergence of the results illustrated below implies that $r_{jk} - \overline{r}_{jk} \xrightarrow[N \to \infty]{} 0$ with $\overline{r}_{jk} = \log_2(1 + \overline{\gamma}_{jk})$, where $\overline{\gamma}_{jk}$ denotes one of the asymptotic approximations computed below.

As limiting cases are considered, the following conditions (widely used in the literature [5], [16], [32], [33]) are needed.

$$\text{A1}: \limsup ||\boldsymbol{\Theta}_{jlk}^{1/2}|| < \infty \quad \text{and} \quad \liminf \frac{1}{N}\text{tr}\left(\boldsymbol{\Theta}_{jlk}\right) > 0$$

$$\text{A2}: \exists \epsilon > 0: \lambda_{\min}\left(\frac{1}{N}\mathbf{H}_{ll}^H \mathbf{H}_{ll}\right) > \epsilon$$

$$\text{A3}: \limsup_{N} ||\frac{1}{N}\mathbf{Z}_l|| < \infty$$

$$\text{A4}: \text{rank}(\hat{\mathbf{H}}_{ll}) \geq K.$$

### A. Large System Results for Vector Normalization

In this subsection, we derive DEs for $\gamma_{jk}^{\text{VN}}$, when any of MRT, ZF, and RZF precoding schemes is used.

$$\gamma_{jk}^{\text{VN}} = \frac{d_{jk} \, |\mathbb{E}[\mathbf{h}_{jjk}^H \mathbf{f}_{jk}]|^2}{\frac{\sigma^2}{\rho_{\text{dl}}} + d_{jk} \, \text{var}(\mathbf{h}_{jjk}^H \mathbf{f}_{jk}) + \sum_{l=1}^{L} \sum_{i=1, i \neq k}^{K} d_{li} \, \mathbb{E}[|\mathbf{h}_{ljk}^H \mathbf{f}_{li}|^2] + \sum_{l=1, l \neq j}^{L} d_{lk} \, \mathbb{E}[|\mathbf{h}_{ljk}^H \mathbf{f}_{lk}|^2]} \tag{16}$$

$$\gamma_{jk}^{\text{MN}} = \frac{\eta_j \, |\mathbb{E}[\mathbf{h}_{jjk}^H \mathbf{f}_{jk}]|^2}{\frac{\sigma^2}{\rho_{\text{dl}}} + \eta_j \, \text{var}(\mathbf{h}_{jjk}^H \mathbf{f}_{jk}) + \sum_{l=1}^{L} \sum_{i=1, i \neq k}^{K} \eta_l \, \mathbb{E}[|\mathbf{h}_{ljk}^H \mathbf{f}_{li}|^2] + \sum_{l=1, l \neq j}^{L} \eta_l \, \mathbb{E}[|\mathbf{h}_{ljk}^H \mathbf{f}_{lk}|^2]}. \tag{17}$$

---

**Theorem 1.** *Let A1 hold true. If MRT with VN is used, then $\gamma_{jk}^{\text{VN}} - \overline{\gamma}_{jk}^{(\text{MRT}-\text{VN})} \xrightarrow[N \to \infty]{} 0$ almost surely with*

$$\overline{\gamma}_{jk}^{(\text{MRT}-\text{VN})} = \frac{d_{jk}^\dagger \left(\frac{1}{N}\text{tr}\mathbf{\Phi}_{jjk}\right)^2}{\frac{\sigma^2}{N\rho_{\text{dl}}} + \frac{1}{N}\sum_{l=1}^{L}\sum_{i=1}^{K} d_{li}^\dagger z_{li,jk} + \sum_{l=1,l\neq j}^{L} d_{lk}^\dagger |\frac{1}{N}\text{tr}\mathbf{\Phi}_{ljk}|^2} \tag{18}$$

*where*

$$d_{li}^\dagger = \left(\frac{1}{N}\text{tr}\mathbf{\Phi}_{lli}\right)^{-1} \tag{19}$$

$$z_{li,jk} = \frac{1}{N}\text{tr}\mathbf{\Theta}_{ljk}\mathbf{\Phi}_{lli}. \tag{20}$$

*Proof.* The proof is provided in Appendix A. □

**Theorem 2.** *Let A1 and A3 hold true. If RZF with VN is used, then $\gamma_{jk}^{\text{VN}} - \overline{\gamma}_{jk}^{(\text{RZF}-\text{VN})} \xrightarrow[N \to \infty]{} 0$ almost surely while*

$$\overline{\gamma}_{jk}^{(\text{RZF}-\text{VN})} = \frac{d_{jk}^\circ \frac{u_{jk}^2}{(1+u_{jk})^2}}{\frac{\sigma^2}{N\rho_{dl}} + \frac{1}{N}\sum_{l=1}^{L}\sum_{i=1}^{K} d_{li}^\circ \frac{\epsilon_{li,jk}}{(1+u_{li})^2} + \sum_{l=1,l\neq j}^{L} d_{lk}^\circ \frac{|u_{ljk}|^2}{(1+u_{lk})^2}} \tag{21}$$

*with*

$$d_{li}^\circ = \frac{(1+u_{li})^2}{\frac{1}{N}\text{tr}\mathbf{\Phi}_{lli}\mathbf{T}'_{l,\mathbf{I}_N}} \tag{22}$$

$$u_{lk} = \frac{1}{N}\text{tr}\mathbf{\Phi}_{llk}\mathbf{T}_l \tag{23}$$

$$u_{ljk} = \frac{1}{N}\text{tr}\mathbf{\Phi}_{ljk}\mathbf{T}_l \tag{24}$$

$$\epsilon_{li,jk} = \frac{1}{N}\text{tr}\mathbf{\Theta}_{ljk}\mathbf{T}'_{l,\mathbf{\Phi}_{lli}} + \frac{|u_{ljk}|^2}{(1+u_{lk})^2} \times \frac{1}{N}\text{tr}\mathbf{\Phi}_{llk}\mathbf{T}'_{l,\mathbf{\Phi}_{lli}}$$
$$- \frac{2}{1+u_{lk}} \text{Re}\left(\frac{1}{N}\text{tr}\mathbf{\Phi}_{ljk}\mathbf{T}'_{l,\mathbf{\Phi}_{lli}} \times u_{ljk}^*\right) \tag{25}$$

*and $\mathbf{S}_l = \frac{\mathbf{Z}_l}{N}$. Also, $\mathbf{T}_l$, $\mathbf{T}'_{l,\mathbf{I}_N}$, and $\mathbf{T}'_{l,\mathbf{\Phi}_{lli}}$ are given in Theorems 7 and 8 in Appendix E.*

*Proof.* The proof is provided in Appendix B. □

**Theorem 3.** *Let A1, A2 and A4 hold true. If ZF with VN is employed, then $\gamma_{jk}^{\text{VN}} - \overline{\gamma}_{jk}^{(\text{ZF}-\text{VN})} \xrightarrow[N \to \infty]{} 0$ almost surely with*

$$\overline{\gamma}_{jk}^{(\text{ZF}-\text{VN})} = \frac{\underline{u}_{jk}}{\frac{\sigma^2}{N\rho_{\text{dl}}} + \frac{1}{N}\sum_{l=1}^{L}\sum_{i=1}^{K} \frac{\underline{\epsilon}_{li,jk}}{\underline{u}_{li}} + \sum_{l=1,l\neq j}^{L} \frac{\underline{u}_{ljk}^2}{\underline{u}_{lk}}} \tag{26}$$

*where*

$$\underline{u}_{li} = \frac{1}{N}\text{tr}\left(\mathbf{\Phi}_{lli}\underline{\mathbf{T}}_l\right) \tag{27}$$

$$\underline{\mathbf{T}}_l = \left(\frac{1}{N}\sum_{i=1}^{K} \frac{\mathbf{\Phi}_{lli}}{\underline{u}_{li}} + \mathbf{I}_N\right)^{-1} \tag{28}$$

$$\underline{u}_{ljk} = \frac{1}{N}\text{tr}\left(\mathbf{\Phi}_{ljk}\underline{\mathbf{T}}_l\right) \tag{29}$$

$$\underline{\epsilon}_{li,jk} = \frac{1}{N}\text{tr}\mathbf{\Theta}_{ljk}\underline{\mathbf{T}}'_{l,\mathbf{\Phi}_{lli}} + \frac{|\underline{u}_{ljk}|^2}{\underline{u}_{lk}^2}\frac{1}{N}\text{tr}\mathbf{\Phi}_{llk}\underline{\mathbf{T}}'_{l,\mathbf{\Phi}_{lli}}$$
$$- \frac{2}{\underline{u}_{lk}} \text{Re}\left(\underline{u}_{ljk}^* \frac{1}{N}\text{tr}\mathbf{\Phi}_{ljk}\underline{\mathbf{T}}'_{l,\mathbf{\Phi}_{lli}}\right) \tag{30}$$

$$\underline{\mathbf{T}}'_{l,\mathbf{\Phi}_{llk}} = \underline{\mathbf{T}}_l \left(\frac{1}{N}\sum_{i=1}^{K} \frac{\underline{u}'_{li,\mathbf{\Phi}_{llk}}\mathbf{\Phi}_{lli}}{\underline{u}_{li}^2} + \mathbf{\Phi}_{llk}\right) \underline{\mathbf{T}}_l \tag{31}$$

*where $\underline{\mathbf{u}}'_{l,\mathbf{\Phi}_{llk}} = [\underline{u}'_{l1,\mathbf{\Phi}_{llk}}, \ldots, \underline{u}'_{lK,\mathbf{\Phi}_{llk}}]^T \in \mathbb{C}^K$ is computed as*

$$\underline{\mathbf{u}}'_{l,\mathbf{\Phi}_{llk}} = (\mathbf{I}_K - \underline{\mathbf{J}}_l)^{-1} \underline{\mathbf{v}}_{l,\mathbf{\Phi}_{llk}} \tag{32}$$

*with the entries of $\underline{\mathbf{J}}_l \in \mathbb{C}^{K \times K}$ and $\underline{\mathbf{v}}_{l,\mathbf{\Phi}_{llk}} \in \mathbb{C}^K$ are given by:*

$$[\underline{\mathbf{J}}_l]_{n,i} = \frac{1}{N^2} \frac{\text{tr}\left(\mathbf{\Phi}_{lln}\underline{\mathbf{T}}_l\mathbf{\Phi}_{lli}\underline{\mathbf{T}}_l\right)}{\underline{u}_{li}^2} \tag{33}$$

$$[\underline{\mathbf{v}}_{l,\mathbf{\Phi}_{llk}}]_i = \frac{1}{N}\text{tr}\left(\mathbf{\Phi}_{lli}\underline{\mathbf{T}}_l\mathbf{\Phi}_{llk}\underline{\mathbf{T}}_l\right). \tag{34}$$

*Proof.* The proof is provided in the Appendix C. □

Notice that the computation of the DEs with ZF precoding (either VN or MN) for the considered multicell Massive MIMO system is more involved than with MRT or RZF precoding schemes. This is mainly due to the fact that it is not straightforward to start with ZF precoder and then compute the DEs by applying common techniques, e.g., matrix inversion lemma. Therefore, in proving Theorem 3 (and also Theorem 6) we start with the DE of RZF and then use a bounding and limiting technique to compute the DE for ZF.

### B. Large System Results for Matrix Normalization

Next, the DEs of $\gamma_{jk}^{\text{MN}}$ are given for MRT, ZF, and RZF. Note that the DEs of $\gamma_{jk}^{\text{MN}}$ for MRT and RZF are obtained from [5].

**Theorem 4.** *[5, Theorem 4] Let A1 hold true. If MRT with MN is used, then $\gamma_{jk}^{\text{MN}} - \overline{\gamma}_{jk}^{(\text{MRT-MN})} \xrightarrow[N\to\infty]{} 0$ almost surely with*

$$\overline{\gamma}_{jk}^{(\text{MRT-MN})} \triangleq \frac{\lambda_j \left(\frac{1}{N}\text{tr}\boldsymbol{\Phi}_{jjk}\right)^2}{\frac{\sigma^2}{N\rho_{\text{dl}}} + \frac{1}{N}\sum_{l=1}^{L}\sum_{i=1}^{K}\lambda_l z_{li,jk} + \sum_{l=1,l\neq j}^{L}\lambda_l |\frac{1}{N}\text{tr}\boldsymbol{\Phi}_{ljk}|^2} \quad (35)$$

*where $z_{li,jk}$ is given in (20) and*

$$\lambda_j = \left(\frac{1}{K}\sum_{k=1}^{K}\frac{1}{N}\text{tr}\boldsymbol{\Phi}_{jjk}\right)^{-1}. \quad (36)$$

**Theorem 5.** *[5, Theorem 6] Let A1 and A3 hold true. If RZF with MN is used, then $\gamma_{jk}^{\text{MN}} - \overline{\gamma}_{jk}^{(\text{RZF-MN})} \xrightarrow[N\to\infty]{} 0$ almost surely with*

$$\overline{\gamma}_{jk}^{(\text{RZF-MN})} \triangleq \frac{\lambda_j \frac{u_{jk}^2}{(1+u_{jk})^2}}{\frac{\sigma^2}{N\rho_{\text{dl}}} + \frac{1}{N}\sum_{l=1}^{L}\sum_{i=1}^{K}\lambda_l \frac{\epsilon_{li,jk}}{(1+u_{li})^2} + \sum_{l=1,l\neq j}^{L}\lambda_l \frac{|u_{ljk}|^2}{(1+u_{lk})^2}} \quad (37)$$

*with*

$$\lambda_l = \frac{K}{N}\left(\frac{1}{N}\text{tr}\mathbf{T}_l - \frac{1}{N}\text{tr}(\frac{\mathbf{Z}_l}{N} + \alpha_l \mathbf{I}_N)\mathbf{T}'_{l,\mathbf{I}_N}\right)^{-1} \quad (38)$$

*where $\mathbf{S}_l = \frac{\mathbf{Z}_l}{N}$ and $\mathbf{T}_l$ and $\mathbf{T}'_{l,\mathbf{I}_N}$ are given by Theorem 7 and Theorem 8. Also $u_{li}$, $u_{ljk}$, and $\epsilon_{li,jk}$ are defined in Theorem 2.*

**Theorem 6.** *Let A1, A2 and A4 hold true. If ZF with MN is used, then $\gamma_{jk}^{\text{MN}} - \overline{\gamma}_{jk}^{(\text{ZF-MN})} \xrightarrow[N\to\infty]{} 0$ almost surely with*

$$\overline{\gamma}_{jk}^{(\text{ZF-MN})} = \frac{\lambda_j}{\frac{\sigma^2}{N\rho_{\text{dl}}} + \frac{1}{N}\sum_{l=1}^{L}\sum_{i=1}^{K}\lambda_l \frac{\epsilon_{li,jk}}{u_{li}^2} + \sum_{l=1,l\neq j}^{L}\lambda_l \frac{u_{ljk}^2}{u_{lk}^2}} \quad (39)$$

*with $\lambda_j = \left(\frac{1}{K}\sum_{i=1}^{K}\frac{1}{u_{ji}}\right)^{-1}$ where $u_{li}$, $u_{ljk}$, and $\epsilon_{li,jk}$ are given in Theorem 3.*

*Proof sketch.* The proof follows the same procedure as the proof of Theorem 3 presented in Appendix C. Start with the triangle equality and bound $|\gamma_{lk}^{(\text{ZF-MN})} - \overline{\gamma}_{lk}^{(\text{ZF-MN})}|$. Then find the DE of $\overline{\gamma}_{jk}^{(\text{ZF-MN})}$ by letting $\alpha \to 0$ in $\overline{\gamma}_{jk}^{(\text{RZF-MN})}$. □

The asymptotic expressions provided in Theorems 1, 2, 3, and 6 will be shown to be very tight, even for systems with finite dimensions, by means of numerical results in Section V. This allows us to use them for evaluating the performance of practical Massive MIMO systems without the need for time-consuming Monte Carlo simulations. Moreover, they lay the foundation for further analysis of different configurations of Massive MIMO systems (e.g., distributed massive MIMO systems [34], [35]). Next, they are used to get further insights into the system under investigation for uncorrelated fading channels.

## IV. Effect of Power Normalization Techniques

In this section, we use the asymptotic approximations provided above to gain novel insights into the interplay between the different system parameters and the power normalization techniques in Massive MIMO. To this end, we consider a special case of the general channel model of (1) in which $\boldsymbol{\Theta}_{jlk} = d_{jlk}\mathbf{I}_N$ such that

$$\mathbf{h}_{jlk} = \sqrt{d_{jlk}}\mathbf{z}_{jlk} \quad (40)$$

where $\mathbf{z}_{jlk} \sim \mathcal{CN}(\mathbf{0}, \mathbf{I}_N)$ and $d_{jlk}$ accounts for an arbitrary large-scale fading coefficient including pathloss and shadowing. Note this corresponds to a uncorrelated fading channel model, which is a quite popular model in Massive MIMO that allows us to capture the essence of the technology [1], [6]. Under the above circumstances, we have that:

**Corollary 1.** *Let $\lambda_j = \bar{u}\left(\frac{1}{K}\sum_{i=1}^{K}\frac{\alpha_{ji}}{d_{jji}^2}\right)^{-1}$ and $\bar{u} = 1 - \frac{K}{N}$. If the channel is modelled as in (40), then*

$$\overline{\gamma}_{jk}^{(\text{ZF-VN})} = \frac{\frac{d_{jjk}^2}{\alpha_{jk}}\bar{u}}{\nu_{jk} + \underbrace{\sum_{l=1,l\neq j}^{L}\bar{u}\frac{d_{ljk}^2}{\alpha_{lk}}}_{\text{Pilot Contamination}}} \quad (41)$$

$$\overline{\gamma}_{jk}^{(\text{ZF-MN})} = \frac{\lambda_j}{\nu_{jk} + \underbrace{\sum_{l=1,l\neq j}^{L}\lambda_l \frac{d_{ljk}^2}{d_{llk}^2}}_{\text{Pilot Contamination}}} \quad (42)$$

*where*

$$\nu_{jk} = \underbrace{\frac{\sigma^2}{N\rho_{\text{dl}}}}_{\text{Noise}} + \underbrace{\frac{K}{N}\sum_{l=1}^{L}d_{ljk}\left(1 - \frac{d_{ljk}}{\alpha_{lk}}\right)}_{\text{Interference}}. \quad (43)$$

*with $\alpha_{lk} = \sum_{n=1}^{L}d_{lnk} + \frac{\sigma^2}{\rho_{\text{tr}}}$.*

*Proof.* See Appendix D. □

**Corollary 2.** *Let $\theta_l = \left(\frac{1}{K}\sum_{i=1}^{K}\frac{d_{lli}^2}{\alpha_{li}}\right)^{-1}$. If the channel is modelled as in (40) and MRT is used, then*

$$\overline{\gamma}_{jk}^{(\text{MRT-VN})} = \frac{\frac{d_{jjk}^2}{\alpha_{jk}}}{\vartheta_{jk} + \underbrace{\sum_{l=1,l\neq j}^{L}\frac{d_{ljk}^2}{\alpha_{lk}}}_{\text{Pilot Contamination}}} \quad (44)$$

$$\overline{\gamma}_{jk}^{(\text{MRT-MN})} = \frac{\theta_j \left(\frac{d_{jjk}^2}{\alpha_{jk}}\right)^2}{\vartheta_{jk} + \underbrace{\sum_{l=1,l\neq j}^{L}\theta_l \left(\frac{d_{llk}d_{ljk}}{\alpha_{lk}}\right)^2}_{\text{Pilot Contamination}}} \quad (45)$$

*with*

$$\vartheta_{jk} = \underbrace{\frac{\sigma^2}{N\rho_{\text{dl}}}}_{\text{Noise}} + \underbrace{\frac{K}{N}\sum_{l=1}^{L}d_{ljk}}_{\text{Interference}}. \quad (46)$$

*Proof.* The proof follows a similar procedure as that of Corollary 1. □

The results of Corollaries 1 and 2 are instrumental in obtaining the following insights into MRT and ZF with either MN or VN.

**Remark 1** (Effect of VN and MN)**.** The terms $\nu_{jk}$ and $\vartheta_{jk}$ in (43) and (46) are the same for both VN and MN. This means that both normalization techniques have exactly the same effect on the resulting noise and interference terms experienced by each UE in the system. On the other hand, they affect differently the signal and pilot contamination powers. The expressions (41)-(46) explicitly state the relation between the SINR contributions (signal, interference, noise, and pilot contamination), the propagation environment, and the two normalization techniques for ZF and MRT precoding schemes.

**Remark 2** (On the mutual effect of UEs)**.** If VN is employed, then the signal power and the pilot contamination of UE $k$ in cell $j$, for both MRT and ZF precoding, depends only on the coefficients $d_{lnk}$ $\forall l, n \in \mathcal{L}$ through $\alpha_{lk}$. This means that they are both affected only by the large-scale gains of the UEs in the network using the same pilot. On the other hand, under MN both terms depend on the coefficients $\lambda_l$ $\forall l \in \mathcal{L}$ (or $\theta_l$ for MRT) and thus are influenced by all the UEs in the network, even though they make use of different pilot sequences.

**Remark 3** (Large-scale fading and power normalization)**.** Assume that the large-scale fading is neglected such that it is the same for every UE in the network, i.e., $d_{ljk} = d$ $\forall l, j, k$. Then, the expressions in (41) and (42) for ZF and those in (44) and (45) for MRT become equal. This means that the large-scale fading has a fundamental impact on VN and MN and cannot be ignored.

Consider now, for further simplicity, a single-cell setup, i.e., $L = 1$. Dropping the cell index, $\alpha_{lk}$ reduces to $\alpha_k = d_k + \sigma^2/\rho_{\text{tr}}$. Also assume that the UEs operate in the high training SNR regime such that $\rho_{\text{tr}} \gg 1$. Under these conditions, we have that:

**Lemma 1.** *If $L = 1$ and $\rho_{\text{tr}} \gg 1$, then for ZF precoding, VN outperforms MN in terms of sum rate and the sum rate gap $\Delta r \geq 0$ is given by*

$$\Delta r = \sum_{k=1}^{K} \log\left(1 + \frac{1}{\frac{\sigma^2}{N\rho_{\text{dl}}\bar{u}}\frac{1}{d_k}}\right) - K \log\left(1 + \frac{1}{\frac{\sigma^2}{N\rho_{\text{dl}}\bar{u}}\frac{1}{K}\sum_{i=1}^{K}\frac{1}{d_i}}\right). \quad (47)$$

*Proof.* From Corollary 1, setting $L = 1$ and assuming $\rho_{\text{tr}} \gg 1$ we obtain that $\alpha_k \simeq d_k$ and $\nu_k \simeq \frac{\sigma^2}{N\rho_{\text{dl}}}$. Then, the result follows by applying the Jensen's inequality (by the convexity of $\log(1 + 1/x)$). □

Notice that Lemma 1 extends the results of [26] and [27] to a system that accounts for CSI acquisition and arbitrary pathloss and UEs' distribution. Also, observe that (41) and (42) simplify as:

$$\bar{\gamma}_{jk}^{(\text{ZF-VN})} = \frac{(N-K)\rho_{\text{dl}}}{\sigma^2}d_k \quad (48)$$

$$\bar{\gamma}_{jk}^{(\text{ZF-MN})} = \frac{(N-K)\rho_{\text{dl}}}{\sigma^2}\left(\frac{1}{K}\sum_{i=1}^{K}\frac{1}{d_i}\right)^{-1} \quad (49)$$

from which it follows that VN provides higher SINR to the UEs that are closer to the BS and lower SINR for those that are far away from the BS (which resembles opportunistic resource allocation). On the other hand, MN provides a uniform quality of experience to all UEs. This proves evidence of the fact that ZF with VN resembles a sum rate maximizer. On the other hand, it provides a notion of fairness under MN. Notice that fairness means similar SINR (quality of experience) and it should not be confused with equal power allocation. The above results and observations will be validated below in Section V by means of numerical results. Also, the DEs provided in Corollaries 1 and 2 will be used to investigate the main limiting factors of Massive MIMO.

## V. NUMERICAL RESULTS

Monte-Carlo simulations are now used to validate the asymptotic analysis for different values of $N$ and $K$. We consider a multicell network composed of $L = 7$ cells, one in the center and six around. Each cell radius is 1000 meters. A 20 MHz channel is considered and the thermal noise power is assumed to be $-174$ dBm/Hz. The UEs are randomly and uniformly distributed within each cell excluding a circle of radius 100 meters. The channel is modeled as in [36]. In particular, we assume that the matrices $\boldsymbol{\Theta}_{ljk}^{1/2}$ are given by

$$\boldsymbol{\Theta}_{ljk}^{1/2} = \sqrt{d_{ljk}}\mathbf{A} \quad (50)$$

where $\mathbf{A} = [\mathbf{a}(\theta_1), \ldots, \mathbf{a}(\theta_N)] \in \mathbb{C}^N$ with $\mathbf{a}(\theta_i)$ given by

$$\mathbf{a}(\theta_i) = \frac{1}{\sqrt{N}}[1, e^{-i2\pi\omega\sin(\theta_i)}, \ldots, e^{-i2\pi\omega(N-1)\sin(\theta_i)}]^T \quad (51)$$

where $\omega = 0.3$ is the antenna spacing and $\theta_i = -\pi/2 + (i-1)\pi/N$. Also, $d_{ljk}$ is the large-scale attenuation, which is modeled as $d_{ljk} = x_{ljk}^{-\beta}$ where $x_{ljk}$ denotes the distance of UE $k$ in cell $j$ from BS $l$ and $\beta = 3.7$ is the path-loss exponent. We let $\rho_{\text{tr}} = 6$ dB and $\rho_{\text{dl}} = 10$ dB, which corresponds to a practical setting [5]. The results are obtained for 100 different channel and UE distributions realizations.

Figs. 1 and 2 validate the accuracy of the DEs provided in Theorems 1, 2, 3, and 6. In particular, both figures report the ergodic achievable sum rate of the center cell versus $N$ for $K = 8$ and 16, respectively. The solid lines correspond to the asymptotic sum rate whereas the markers are achieved through Monte Carlo simulation. As it is depicted, the asymptotic approximation match perfectly with numerical results. Notice that Figs. 1 and 2 (and also Table 1) extend the results in [26] and [27] in the sense that account for CSI acquisition, pilot contamination, arbitrary pathloss and UEs' distribution.

In Lemma 1, it is shown that ZF under VN conveys a notion of sum rate maximization, while ZF with MN resembles a fairness provisioning precoder. Now, we use Table I to validate this observation and also to verify the accuracy of the computed DEs for the simplified channel model in (40). The first column of Table I reports the number of antennas, the second one is the UE index. The third and fourth columns are the asymptotic (as given in (42)) and simulated SINRs of each UE under MN. The corresponding results for VN are reported in the sixth and seventh columns. The fifth and eighth columns report the percentage of the error while estimating a specific UE SINR with the computed DEs. As predicted by Lemma 1, ZF with MN provides a more uniform experience for all UEs, while ZF with VN provides very high SINRs to specific UEs (UEs 2, 4, and 6) and much lower SINRs to others. More precisely, the SINR variance with MN is equal to 0.8 (5.79) for $N = 40$ ($N = 80$), while for VN it is equal to 2627 (11550 for $N = 80$). Notice also that the percentage of error is always less than $4\%$, which proves the high accuracy of the DEs. Therefore, one can simply use the DEs to achieve insight into the network performance, instead of using time-consuming Monte Carlo simulations. Moreover, the DEs do not contain any randomness and are purely based on the large-scale statistics of the network. Hence, they can be used for network optimization purposes.

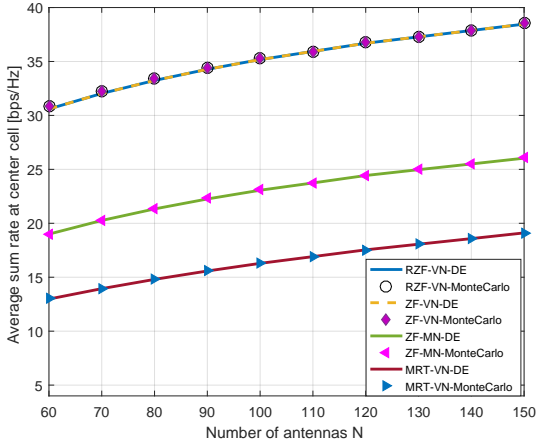

Fig. 1: Ergodic achievable sum rate of center cell for MRT, ZF, and RZF with VN and ZF with MN for $K = 8$.

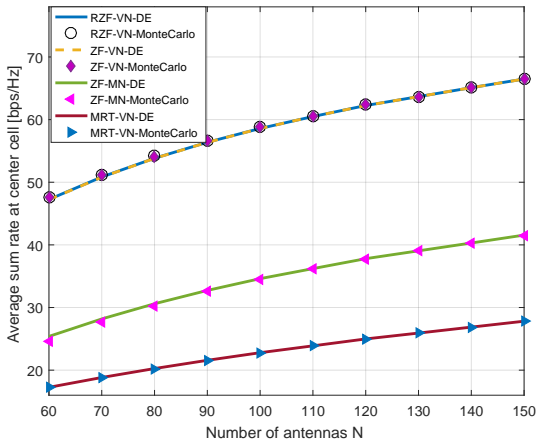

Fig. 2: Ergodic achievable sum rate of center cell for MRT, ZF, and RZF with VN and ZF with MN for $K = 16$.

The DEs given in Corollaries 1 and 2 and Theorems 2

TABLE I: SINR of each UE under ZF with VN and MN.

| No. Ant. | UE | MN DE | MN MC | Er. % | VN DE | VN MC | Er. % |
|---|---|---|---|---|---|---|---|
| $N = 40$ | 1 | 2.25 | 2.19 | 2.6 | 1.85 | 1.84 | 0.5 |
| | 2 | 4.89 | 4.84 | 1.0 | 147 | 149 | 1.3 |
| | 3 | 3.34 | 3.29 | 1.5 | 3.61 | 3.53 | 2.2 |
| | 4 | 5.14 | 5.12 | 0.3 | 37.5 | 37.6 | 0.2 |
| | 5 | 4.09 | 4.02 | 1.7 | 1.97 | 1.96 | 0.5 |
| | 6 | 4.26 | 4.41 | 3.5 | 85 | 87 | 2.3 |
| | 7 | 3.30 | 3.33 | 0.9 | 2.14 | 2.2 | 2.8 |
| | 8 | 3.52 | 3.50 | 0.5 | 2.52 | 2.49 | 1.2 |
| $N = 80$ | 1 | 3.20 | 3.15 | 1.5 | 3.1 | 2.98 | 3.8 |
| | 2 | 10.57 | 10.40 | 1.6 | 316 | 310 | 1.9 |
| | 3 | 5.94 | 6.03 | 1.5 | 6.40 | 6.36 | 0.6 |
| | 4 | 9.33 | 9.47 | 1.5 | 72.2 | 72.2 | 0 |
| | 5 | 8.41 | 8.62 | 2.5 | 3.87 | 3.90 | 0.7 |
| | 6 | 9.05 | 9.12 | 0.7 | 182 | 185 | 1.6 |
| | 7 | 5.02 | 4.90 | 2 | 3.50 | 3.43 | 2 |
| | 8 | 5.89 | 5.74 | 2.5 | 4.28 | 4.21 | 1.6 |

and 5 are now used to investigate a common belief in the Massive MIMO literature, that is: under uncorrelated fading when $N \to \infty$ the noise and interference contributions vanish asymptotically and pilot contamination becomes the unique bottleneck of the system performance. This follows also from the results in Corollaries 1 and 2 by letting $N$ grow large with $K$ kept fixed. However, in [29] it is shown that it is desirable for Massive MIMO systems to work in a regime where $\frac{N}{K} \leq 10$. Therefore, it is interesting to see what is the major impairment for Massive MIMO under this practical regime: ($i$) is it pilot contamination (or coherent interference)?; ($ii$) is it the noise and interference (or more exactly the non-coherent interference)?; ($iii$) how is the answer related to the choice of the power normalization technique and precoding scheme?

To answer these questions, we employ the so-called pilot contamination-to-interference-plus-noise ratio (PCINR) metric, which is computed by using the DEs provided in Corollaries 1 and 2. Fig. 3 plots the PCINR as a function of $N/K$, i.e., the number of degrees of freedom per-user in the system. Although, the optimal operating regime for maximal spectral efficiency is for $N/K < 10$ [29], we consider $N/K$ up to 20 to cover a wider range of Massive MIMO configurations. Moreover, as the interference increases by having more UEs in the system, we consider three different scenarios with $K = 5$, $K = 10$, and $K = 15$.

Fig. 3 is divided into 3 regions based on the significance of the PCINR term such that, as we move away from region 1 towards region 3, the importance of pilot contamination increases while that of the interference plus noise reduces. Region 1 is where the noise and interference are the dominant limiting factors and pilot contamination has a negligible effect—less than $10\%$ of the noise and interference. As it is depicted, MRT with MN operates within this regime, therefore pilot contamination is never a bottleneck for this scheme, which is mainly limited by noise and interference. Notice that by adding more UEs in the system, the PCINR reduces and pilot contamination becomes even less important. Hence, when MRT with MN is studied in Massive MIMO the effect of pilot contamination can be safely neglected.

Region 2 represents the regime where the noise and inter-ference are the main limiting factors, but pilot contamination is not negligible any more. It is interesting to observe that for the other schemes (other than MRT-MN), Massive MIMO often operates within this region. This shows that, although pilot contamination is a major challenge in Massive MIMO, the interference and noise have still the leading role in limiting the system performance.

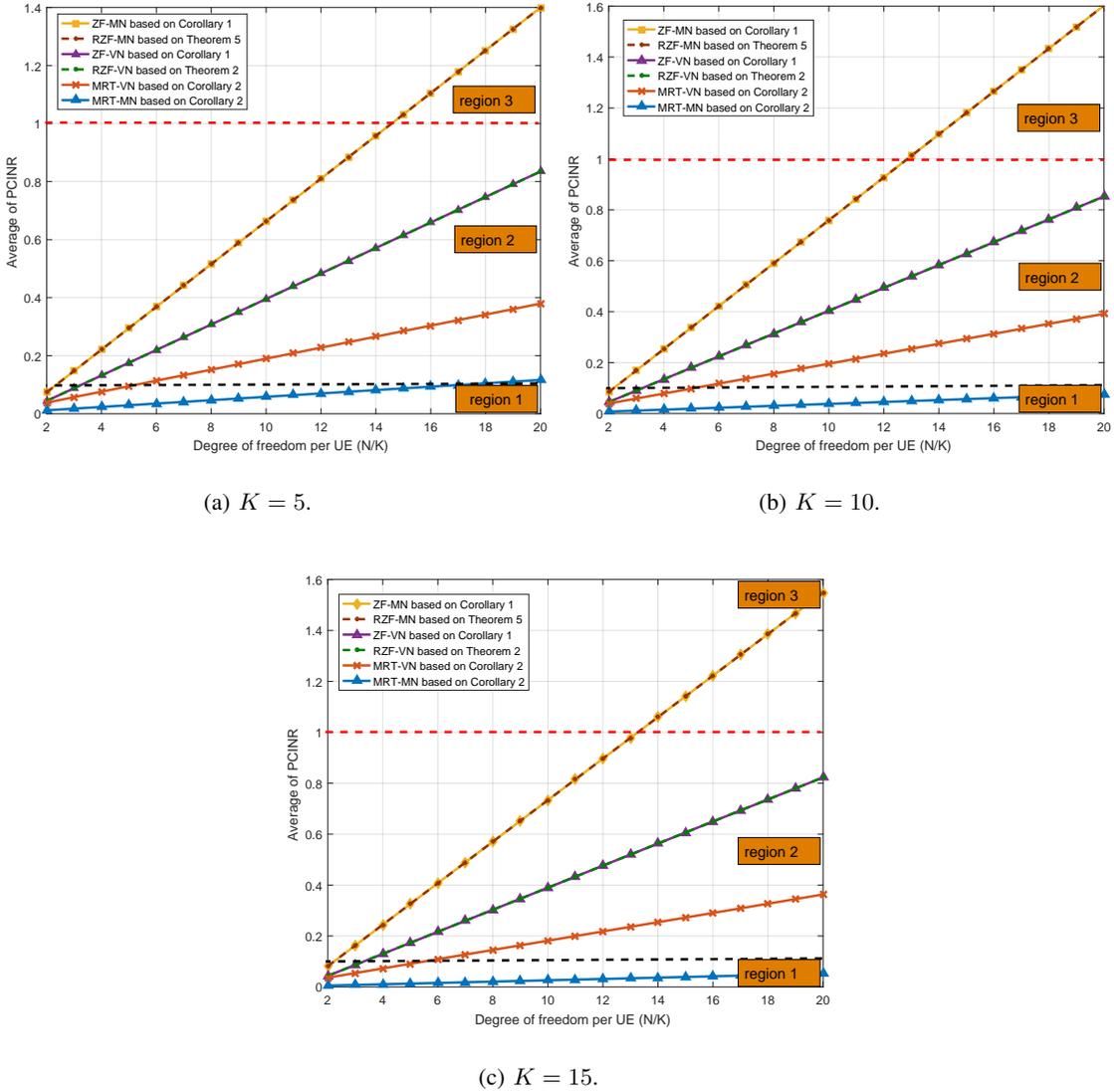

Fig. 3: PCINR versus the degree of freedom per-user, for different values of $K$.

Finally, region 3 presents the superiority of pilot contamination effect. If $K = 10$, then Fig. 3b shows the superiority of interference and noise over pilot contamination for ZF-MN and RZF-MN (ZF-VN and RZF-VN) up to $N = 130$ ($N = 233$) antennas at the BS. With MRT-VN, the system requires more than $N = 510$ to experience the superiority of pilot contamination over interference and noise. This increases to $N = 2650$ with MRT-MN. From Fig. 3, we see also that, for a given value of $N/K$, the value of PCINR for the considered schemes can be ordered as: ZF-MN = RZF-MN $\geq$ ZF-VN = RZF-VN $\geq$ MRT-VN $\geq$ MRT-MN. Based on the above discussion, it is clear that the choice of precoding scheme and normalization technique change the importance of pilot contamination, interference, and noise dramatically and it should be considered carefully when designing Massive MIMO systems.

## VI. CONCLUSIONS

Linear precoding schemes, such as MRT and ZF, have a fundamental role in Massive MIMO. Although these precoding schemes can be employed with optimized power control policies, they are usually implemented by simple matrix or vector power normalization techniques. This is due to the complexity of attaining optimal power control policies [24], as it requires coordination and cooperation among cells and computationally demanding algorithms. On the other hand, the two above precoding power normalization techniques are simple and efficient [2], [3], [5], [6].

This work made use of large system analysis to compute tight asymptotic approximations of the SINR experienced by each UE in the system when using MN or VN. These results can be used to evaluate the performance of practical Massive MIMO systems without the need for time-consuming Monte Carlo simulations. Under uncorrelated fading channels, we analytically showed that MN and VN treat the noise and interference in the same manner, but have different effects on pilot contamination and received signal power. We also revealed the key role played by large-scale fading, positions of UEs, and pilot assignment into power normalization. We explained how a simple change in power normalization can resemble two totally different behaviors, namely, sum-rate maximization or fairness provisioning. Moreover, we showed numerically how the choice of the normalization technique can change the main bottleneck of Massive MIMO systems.

## APPENDIX A

We begin by plugging (7) and (10) into $\gamma_{jk}^{\text{VN}}$ given in (16) to obtain $\gamma_{jk}^{(\text{MRT}-\text{VN})}$. Then, we divide the numerator and denominator of $\gamma_{jk}^{(\text{MRT}-\text{VN})}$ by $N$ and define $d_{li}^\dagger = N d_{li}$. By applying the continuous mapping theorem [31] and replacing each component of $\gamma_{jk}^{(\text{MRT}-\text{VN})}$ by its DE, the DE of $\gamma_{jk}^{(\text{MRT}-\text{VN})}$ is obtained. Notice that DEs of signal power component, variance component, and interference components are given in [5]. Therefore, we only need to compute a DE of the VN coefficient. The latter is given by

$$d_{li} = \frac{1}{\mathbb{E}[\hat{\mathbf{h}}_{lli}^H \hat{\mathbf{h}}_{lli}]} = \frac{1}{\text{tr}\mathbb{E}[\hat{\mathbf{h}}_{lli}\hat{\mathbf{h}}_{lli}^H]} \stackrel{(a)}{=} \frac{1}{\text{tr}\boldsymbol{\Phi}_{lli}} \quad (52)$$

where in $(a)$ we have applied Lemma 3 in Appendix E and used the fact that $\hat{\mathbf{h}}_{lli} \sim \mathcal{CN}(\mathbf{0}, \boldsymbol{\Phi}_{lli})$.

## APPENDIX B

Let us define for convenience

$$\hat{\mathbf{H}}_{ll[k]}\hat{\mathbf{H}}_{ll[k]}^H = \hat{\mathbf{H}}_{ll}\hat{\mathbf{H}}_{ll}^H - \hat{\mathbf{h}}_{llk}\hat{\mathbf{h}}_{llk}^H \quad (53)$$

and $\overline{\mathbf{C}}_l = N\mathbf{C}_l$ and $\overline{\mathbf{C}}_{l[k]} = N\mathbf{C}_{l[k]}$ with

$$\mathbf{C}_l = \left(\hat{\mathbf{H}}_{ll}\hat{\mathbf{H}}_{ll}^H + \mathbf{Z}_l + \alpha_l N \mathbf{I}_N\right)^{-1} \quad (54)$$

$$\mathbf{C}_{l[k]} = \left(\hat{\mathbf{H}}_{ll[k]}\hat{\mathbf{H}}_{ll[k]}^H + \mathbf{Z}_l + \alpha_l N \mathbf{I}_N\right)^{-1}. \quad (55)$$

Plugging (9) and (10) into (16) yields $\gamma_{jk}^{(\text{RZF}-\text{VN})}$. Then, we divide the numerator and denominator by $N$ and replace each term in $\gamma_{jk}^{(\text{RZF}-\text{VN})}$ with its DE. Notice that DEs of signal power component, variance component, and interference components are given in [5]. Therefore, we only need to compute a DE of the VN coefficient. From (10), we have that

$$d_{li}^\circ = \frac{1}{N}d_{li} = \frac{1}{N}(\mathbb{E}[\hat{\mathbf{h}}_{lli}^H \mathbf{C}_l^2 \hat{\mathbf{h}}_{lli}])^{-1}. \quad (56)$$

An asymptotic approximation for $\hat{\mathbf{h}}_{lli}^H \mathbf{C}_l^2 \hat{\mathbf{h}}_{lli}$ can be computed as follows

$$\hat{\mathbf{h}}_{lli}^H \mathbf{C}_l^2 \hat{\mathbf{h}}_{lli} \stackrel{a}{=} \frac{\hat{\mathbf{h}}_{lli}^H \mathbf{C}_{l[i]}^2 \hat{\mathbf{h}}_{lli}}{(1+\hat{\mathbf{h}}_{lli}^H \mathbf{C}_{l[i]} \hat{\mathbf{h}}_{lli})^2} \stackrel{b}{\asymp} \frac{\frac{1}{N^2}\text{tr}\boldsymbol{\Phi}_{lli}\overline{\mathbf{C}}_l^2}{(1+\frac{1}{N}\text{tr}\boldsymbol{\Phi}_{lli}\overline{\mathbf{C}}_l)^2}$$

$$\stackrel{c}{\asymp} \frac{\frac{1}{N^2}\text{tr}\boldsymbol{\Phi}_{lli}\mathbf{T}'_{l,\mathbf{I}_N}}{(1+\frac{1}{N}\text{tr}\boldsymbol{\Phi}_{lli}\mathbf{T}_l^{-1})^2} \quad (57)$$

where (a) follows from Lemma 2 in Appendix E, (b)[1] is obtained by applying Lemmas 3 and 4, and (c) follows from Theorems 7 and 8 with $\mathbf{S}_l = \frac{\mathbf{Z}_l}{N}$. By applying the continuous mapping theorem and the dominated convergence theorem yields

$$d_{li}^\circ \asymp \frac{(1+u_{li})^2}{\frac{1}{N}\text{tr}\boldsymbol{\Phi}_{lli}\mathbf{T}'_{l,\mathbf{I}_N}}. \quad (58)$$

## APPENDIX C

The main idea is to first compute a DE for RZF with $\alpha_l = \alpha$ $\forall l \in \mathcal{L}$ and then to obtain a DE for ZF by letting $\alpha \to 0$. By using the triangle inequality $|\gamma_{lk}^{(\text{ZF}-\text{VN})} - \overline{\gamma}_{lk}^{(\text{ZF}-\text{VN})}|$ can be bounded as follows

$$|\gamma_{lk}^{(\text{ZF}-\text{VN})} - \overline{\gamma}_{lk}^{(\text{ZF}-\text{VN})}| \leq |\gamma_{lk}^{(\text{ZF}-\text{VN})} - \gamma_{lk}^{(\text{RZF}-\text{VN})}| +$$
$$+ |\gamma_{lk}^{(\text{RZF}-\text{VN})} - \overline{\gamma}_{lk}^{(\text{RZF}-\text{VN})}| + |\overline{\gamma}_{lk}^{(\text{RZF}-\text{VN})} - \overline{\gamma}_{lk}^{(\text{ZF}-\text{VN})}|. \quad (59)$$

Next, we show that each term in the right hand side of (59) can be made arbitrarily small (i.e. smaller than any given $\varepsilon > 0$) when $\alpha \to 0$. Let us start with $|\gamma_{lk}^{(\text{ZF}-\text{VN})} - \gamma_{lk}^{(\text{RZF}-\text{VN})}|$. Notice that $\gamma_{lk}^{(\text{ZF}-\text{VN})}$ and $\gamma_{lk}^{(\text{RZF}-\text{VN})}$ are different because of the different form of $\mathbf{F}_l$ in (8) and (9). As $\alpha \to 0$ and for $\mathbf{Z}_l = \mathbf{0}$, we have

$$\lim_{\mathbf{Z}_l = \mathbf{0}, \alpha \to 0} \mathbf{f}_{lk}^{\text{RZF}} = \lim_{\alpha \to 0} \left(\hat{\mathbf{H}}_{ll}\hat{\mathbf{H}}_{ll}^H + N\alpha \mathbf{I_N}\right)^{-1} \hat{\mathbf{H}}_{ll}\mathbf{e}_k$$
$$= \hat{\mathbf{H}}_{ll}\left(\hat{\mathbf{H}}_{ll}^H \hat{\mathbf{H}}_{ll}\right)^{-1} \mathbf{e}_k = \mathbf{f}_{lk}^{\text{ZF}}. \quad (60)$$

Therefore, the term $|\gamma_{lk}^{(\text{ZF}-\text{VN})} - \gamma_{lk}^{(\text{RZF}-\text{VN})}|$ can be made arbitrarily small $\alpha \to 0$. By applying Theorem 2, we have that the second term is such that $|\gamma_{lk}^{(\text{RZF}-\text{VN})} - \overline{\gamma}_{lk}^{(\text{RZF}-\text{VN})}| \asymp 0$ for any $\alpha > 0$. Consider now the third term $|\overline{\gamma}_{lk}^{(\text{RZF}-\text{VN})} - \overline{\gamma}_{lk}^{(\text{ZF}-\text{VN})}|$. Let us define $\overline{\gamma}_{lk}^{(\text{ZF}-\text{VN})} = \lim_{\alpha \to 0} \overline{\gamma}_{lk}^{(\text{RZF}-\text{VN})}$. Observe that

$$\lim_{\alpha \to 0} \overline{\gamma}_{jk}^{(\text{RZF}-\text{VN})} \quad (61)$$

$$= \lim_{\alpha \to 0} \frac{d_{jk}^\circ \frac{u_{jk}^2}{(1+u_{jk})^2}}{\frac{\sigma^2}{N\rho_{\text{dl}}} + \frac{1}{N}\sum_{l=1}^{L}\sum_{i=1}^{K} d_{li}^\circ \frac{\epsilon_{li,jk}}{(1+u_{li})^2} + \sum_{l=1,l\neq j}^{L} d_{lk}^\circ \frac{|u_{ljk}|^2}{(1+u_{lk})^2}}$$

$$= \lim_{\alpha \to 0} \frac{d_{jk}^\circ \frac{\alpha^2 u_{jk}^2}{(\alpha+\alpha u_{jk})^2}}{\frac{\sigma^2}{N\rho_{\text{dl}}} + \frac{1}{N}\sum_{l=1}^{L}\sum_{i=1}^{K} d_{li}^\circ \frac{\alpha^2 \epsilon_{li,jk}}{(\alpha+\alpha u_{li})^2} + \sum_{l=1,l\neq j}^{L} d_{lk}^\circ \frac{|\alpha u_{ljk}|^2}{(\alpha+\alpha u_{lk})^2}}.$$

[1] $a_N \asymp b_N$ is equivalent to $a_N - b_N \xrightarrow[N\to\infty]{} 0$.

Define $\underline{u_{lk}} := \lim_{\alpha \to 0} \alpha u_{lk}$ for every $l$ and $k$. Based on [16] and by replacing $u_{lk}$ from Theorem 2 we have

$$\underline{u_{lk}} = \lim_{\alpha \to 0} \alpha u_{lk} = \lim_{\alpha \to 0} \frac{1}{N} \mathrm{tr} \boldsymbol{\Phi}_{llk} \left( \frac{1}{N} \sum_{i=1}^{K} \frac{\boldsymbol{\Phi}_{lli}}{\alpha u_{li}} + \mathbf{I}_N \right)^{-1}$$

$$= \frac{1}{N} \mathrm{tr} \left( \boldsymbol{\Phi}_{llk} \underline{\mathbf{T}_l} \right) \quad (62)$$

with

$$\underline{\mathbf{T}_l} = \left( \frac{1}{N} \sum_{i=1}^{K} \frac{\boldsymbol{\Phi}_{lli}}{\underline{u_{li}}} + \mathbf{I}_N \right)^{-1}. \quad (63)$$

Also defining $\underline{u_{ljk}} \triangleq \lim_{\alpha \to 0} \alpha u_{ljk}$ for every $l$, $j$, and $k$ we have

$$\underline{u_{ljk}} = \lim_{\alpha \to 0} \alpha \frac{1}{N} \mathrm{tr} \boldsymbol{\Phi}_{ljk} \mathbf{T}_l = \frac{1}{N} \mathrm{tr} \boldsymbol{\Phi}_{ljk} \underline{\mathbf{T}_l}. \quad (64)$$

For the term $\lim_{\alpha \to 0} d_{li}^\circ$, we obtain

$$\lim_{\alpha \to 0} d_{li}^\circ = \lim_{\alpha \to 0} \frac{(1+u_{li})^2}{\frac{1}{N} \mathrm{tr} \boldsymbol{\Phi}_{lli} \mathbf{T}'_{l,\mathbf{I}_N}}$$

$$= \lim_{\alpha \to 0} \frac{u_{li}^2}{\alpha^2 \frac{1}{N} \mathrm{tr} \boldsymbol{\Phi}_{lli} \mathbf{T}'_{l,\mathbf{I}_N}} = \lim_{\alpha \to 0} \frac{u_{li}^2}{\alpha^2 u'_{li,\mathbf{I}_N}} \quad (65)$$

where $\mathbf{T}'_{l,\mathbf{I}_N}$ and $u'_{li,\mathbf{I}_N}$ are given in Theorem 8 in Appendix E. Notice that

$$\lim_{\alpha \to 0} \alpha^2 u'_{li,\mathbf{I}_N} = \alpha^2 \frac{1}{N} \mathrm{tr} \boldsymbol{\Phi}_{lli} \mathbf{T}'_{l,\mathbf{I}_N}$$

$$= \lim_{\alpha \to 0} \alpha^2 \frac{1}{N} \mathrm{tr} \boldsymbol{\Phi}_{lli} \mathbf{T}_l \left( \frac{1}{N} \sum_{t=1}^{K} \frac{u'_{lt,\mathbf{I}_N} \boldsymbol{\Phi}_{llt}}{(1+u_{lt})^2} + \mathbf{I}_M \right) \mathbf{T}_l$$

$$= \frac{1}{N} \mathrm{tr} \boldsymbol{\Phi}_{lli} \underline{\mathbf{T}_l} \left( \frac{1}{N} \sum_{t=1}^{K} \frac{(\lim_{\alpha \to 0} \alpha^2 u'_{lt,\mathbf{I}_N}) \boldsymbol{\Phi}_{llt}}{\underline{u_{lt}}^2} + \mathbf{I}_N \right) \underline{\mathbf{T}_l}$$

$$(66)$$

from which, by replacing $\lim_{\alpha \to 0} \alpha^2 u'_{lt,\mathbf{I}_N}$ with $\underline{u_{lt}}$, we have that (66) reduces to $\underline{u_{li}} = \frac{1}{N} \mathrm{tr} \boldsymbol{\Phi}_{lli} \underline{\mathbf{T}_l}$. Therefore, we have that $\lim_{\alpha \to 0} \alpha^2 u'_{lk,\mathbf{I}_N} = \underline{u_{lk}}$. From (65), we can thus conclude that $\underline{d_{li}^\circ} = \underline{u_{li}}$. On the other hand, for $\epsilon_{li,jk}$ we have

$$\underline{\epsilon_{li,jk}} = \lim_{\alpha \to 0} \alpha^2 \epsilon_{li,jk}$$

$$= \frac{1}{N} \mathrm{tr} \boldsymbol{\Theta}_{ljk} \underline{\mathbf{T}'_{l,\boldsymbol{\Phi}_{lli}}} - \frac{2}{\underline{u_{lk}}} \mathrm{Re} \left( \underline{u_{ljk}^*} \frac{1}{N} \mathrm{tr} \boldsymbol{\Phi}_{ljk} \underline{\mathbf{T}'_{l,\boldsymbol{\Phi}_{lli}}} \right)$$

$$+ \frac{|\underline{u_{ljk}}|^2}{\underline{u_{lk}}^2} \frac{1}{N} \mathrm{tr} \boldsymbol{\Phi}_{llk} \underline{\mathbf{T}'_{l,\boldsymbol{\Phi}_{lli}}} \quad (67)$$

where $\underline{\mathbf{T}'_{l,\boldsymbol{\Phi}_{lli}}} \triangleq \lim_{\alpha \to 0} \alpha^2 \mathbf{T}'_{l,\boldsymbol{\Phi}_{lli}}$ is

$$\underline{\mathbf{T}'_{l,\boldsymbol{\Phi}_{lli}}} = \lim_{\alpha \to 0} \alpha^2 \mathbf{T}_l \left[ \frac{1}{N} \sum_{t=1}^{K} \frac{u'_{lt,\boldsymbol{\Phi}_{lli}} \boldsymbol{\Phi}_{llt}}{(1+u_{lt})^2} + \boldsymbol{\Phi}_{lli} \right] \mathbf{T}_l$$

$$= \underline{\mathbf{T}_l} \left[ \frac{1}{N} \sum_{t=1}^{K} \frac{\underline{u'_{lt,\boldsymbol{\Phi}_{lli}}} \boldsymbol{\Phi}_{llt}}{\underline{u_{lt}}^2} + \boldsymbol{\Phi}_{lli} \right] \underline{\mathbf{T}_l} \quad (68)$$

where $\underline{u'_{lt,\boldsymbol{\Phi}_{lli}}} = \lim_{\alpha \to 0} \alpha^2 u'_{lt,\boldsymbol{\Phi}_{lli}}$. From Theorem 8, we have

$$\underline{\mathbf{u}'_{l,\boldsymbol{\Phi}_{lli}}} = \lim_{\alpha \to 0} (\mathbf{I}_K - \mathbf{J}_l)^{-1} \alpha^2 \mathbf{v}_{l,\boldsymbol{\Phi}_{lli}} = (\mathbf{I}_K - \underline{\mathbf{J}_l})^{-1} \underline{\mathbf{v}_{l,\boldsymbol{\Phi}_{lli}}}$$

where $\underline{\mathbf{J}_l}$ and $\underline{\mathbf{v}_{l,\boldsymbol{\Phi}_{lli}}}$ are given by (33) and (34), respectively. Therefore, $\underline{\epsilon_{lk,jn}} = \lim_{\alpha \to 0} \alpha^2 \epsilon_{lk,jn}$ follows (30). Using all the above results in (61) completes the proof.

## APPENDIX D

For brevity we only consider ZF with VN. The same steps can be used for ZF with MN. If the channel is modelled as in (40), then $\boldsymbol{\Theta}_{ljk} = d_{ljk} \mathbf{I}_N$ and

$$\boldsymbol{\Phi}_{ljk} = \frac{d_{llk} d_{ljk}}{\alpha_{lk}} \mathbf{I}_N \quad (69)$$

with $\alpha_{lk} = \sum_{n=1}^{L} d_{lnk} + \frac{\sigma^2}{\rho_{\mathrm{tr}}}$. Plugging (69) into (27) and (28) yields $\underline{u_{lk}} = \frac{d_{llk}^2}{\alpha_{lk}} \frac{1}{N} \mathrm{tr}(\underline{\mathbf{T}_l})$ with

$$\underline{\mathbf{T}_l} = \left( \frac{1}{N} \sum_{i=1}^{K} \frac{1}{\frac{1}{N} \mathrm{tr}(\underline{\mathbf{T}_l})} + 1 \right)^{-1} \mathbf{I}_N. \quad (70)$$

Call $\bar{u} = \frac{1}{N} \mathrm{tr}(\underline{\mathbf{T}_l})$. Therefore, we have that

$$\bar{u} = \frac{1}{N} \mathrm{tr}(\underline{\mathbf{T}_l}) = \left( \frac{K}{N} \frac{1}{\bar{u}} + 1 \right)^{-1}. \quad (71)$$

Solving with respect to $\bar{u}$ yields $\bar{u} = 1 - \frac{K}{N}$. Then, we eventually have that

$$\underline{u_{lk}} = \frac{d_{llk}^2}{\alpha_{lk}} \bar{u} \quad (72)$$

and also $\underline{u_{ljk}} = \frac{d_{llk} d_{ljk}}{\alpha_{lk}} \bar{u}$. Therefore, the pilot contamination term in $\overline{\gamma}_{jk}^{(\mathrm{ZF-VN})}$ reduces to

$$\sum_{l=1,l\neq j}^{L} \frac{\underline{u_{ljk}^2}}{\underline{u_{lk}}} = \sum_{l=1,l\neq j}^{L} \frac{d_{ljk}^2}{\alpha_{lk}} \bar{u}. \quad (73)$$

Let's now compute $[\underline{\mathbf{J}_l}]_{n,i}$ defined as in (33). Using the above results yields

$$[\underline{\mathbf{J}_l}]_{n,i} = \frac{1}{N^2} \frac{d_{lln}^2}{\alpha_{ln}} \frac{d_{lli}^2}{\alpha_{li}} \frac{1}{\underline{u_{li}}^2} \mathrm{tr}(\underline{\mathbf{T}}^2) = \frac{1}{N} \frac{d_{lln}^2}{\alpha_{ln}} \frac{\alpha_{li}}{d_{lli}^2}. \quad (74)$$

Similarly, we have that

$$[\underline{\mathbf{v}_{l,k}}]_i = \frac{d_{lli}^2}{\alpha_{li}} \frac{d_{llk}^2}{\alpha_{lk}} \bar{u}^2. \quad (75)$$

In compact form, we may write $\underline{\mathbf{J}_l}$ and $\underline{\mathbf{v}_{l,k}}$ as

$$\underline{\mathbf{J}_l} = \frac{1}{N} \mathbf{a}_l \mathbf{b}_l^T \qquad \underline{\mathbf{v}_{l,k}} = \frac{d_{llk}^2}{\alpha_{lk}} \bar{u}^2 \mathbf{a}_l \quad (76)$$

with $[\mathbf{a}_l]_i = d_{lli}^2/\alpha_{li}$ and $[\mathbf{b}_l]_i = 1/[\mathbf{a}_l]_i$. Then, we have that (by applying Lemma 2)

$$\underline{\mathbf{u}'_{l,k}} = \frac{d_{llk}^2}{\alpha_{lk}} \bar{u}^2 \left( \mathbf{I}_K - \frac{1}{N} \mathbf{a}_l \mathbf{b}_l^T \right)^{-1} \mathbf{a}_l = \frac{d_{llk}^2}{\alpha_{lk}} \bar{u} \mathbf{a}_l = \underline{u_{lk}} \mathbf{a}_l. \quad (77)$$

Plugging the above result into (31) produces

$$\underline{\mathbf{T}'_{l,\boldsymbol{\Phi}_{lli}}} = \frac{d_{lli}^2}{\alpha_{li}} \underline{\mathbf{T}_l} \left( \frac{K}{N} \frac{1}{\bar{u}} + 1 \right) \underline{\mathbf{T}_l} = \frac{d_{lli}^2}{\alpha_{li}} \bar{u} \mathbf{I}_N = \underline{u_{li}} \mathbf{I}_N. \quad (78)$$

We are thus left with evaluating (30). Using the above results yields

$$\underline{\epsilon_{li,jk}} = \frac{d_{ljk}}{N} \mathrm{tr}(\underline{\mathbf{T}'_{l,\boldsymbol{\Phi}_{lli}}}) - 2 \frac{d_{ljk}}{d_{llk}} \frac{1}{N} \mathrm{tr}(\boldsymbol{\Phi}_{ljk} \underline{\mathbf{T}'_{l,\boldsymbol{\Phi}_{lli}}}) +$$

$$+ \frac{d_{ljk}^2}{d_{llk}^2} \frac{1}{N} \mathrm{tr}(\boldsymbol{\Phi}_{llk} \underline{\mathbf{T}'_{l,\boldsymbol{\Phi}_{lli}}}) \quad (79)$$

from which, using (69) and (78), we obtain

$$\epsilon_{li,jk} = d_{ljk}\underline{u}_{li} - \frac{d_{ljk}^2}{\alpha_{lk}}\underline{u}_{li}.$$

Therefore, we have that

$$\frac{1}{N}\frac{\epsilon_{li,jk}}{\underline{u}_{li}} = \frac{1}{N}d_{ljk}\left(1 - \frac{d_{ljk}}{\alpha_{lk}}\right). \quad (80)$$

Plugging (72), (73) and (80) into (26) produces

$$\frac{1}{N}\sum_{l=1}^{L}\sum_{i=1}^{K}\frac{\epsilon_{li,jk}}{\underline{u}_{li}} = \frac{K}{N}\sum_{l=1}^{L}d_{ljk}\left(1 - \frac{d_{ljk}}{\alpha_{lk}}\right). \quad (81)$$

Collecting all the above results together completes the proof.

## APPENDIX E
## USEFUL RESULTS

**Theorem 7.** *[16, Theorem 1] Let $\mathbf{B}_l = \frac{1}{N}\hat{\mathbf{H}}_{ll}\hat{\mathbf{H}}_{ll}^H + \mathbf{S}_l$ with $\hat{\mathbf{H}}_{ll} \in \mathbb{C}^{N \times K}$ be random with independent column vectors $\hat{\mathbf{h}}_{llk} \sim \mathcal{CN}(\mathbf{0}, \mathbf{\Phi}_{llk})$ for $k \in \{1, \ldots, K\}$, $\mathbf{S}_l \in \mathbb{C}^{N \times N}$ and $\mathbf{Q}_l \in \mathbb{C}^{N \times N}$ be Hermitian nonnegative definite. Assume that $\mathbf{Q}_l$ and the matrices $\mathbf{\Phi}_{llk}$ for $k \in \{1, \ldots, K\}$ have uniformly bounded spectral norms (with respect to $N$). Define*

$$m_{\mathbf{B}_l,\mathbf{Q}_l}(-\alpha_l) \triangleq \frac{1}{N}\mathrm{tr}\mathbf{Q}_l(\mathbf{B}_l + \alpha_l \mathbf{I}_N)^{-1} \quad (82)$$

*Then, for any $\alpha_l > 0$, as $N$ and $K$ grow large with $\beta = \frac{N}{K}$ such that $0 < \liminf \beta \le \limsup \beta < \infty$ we have that*

$$m_{\mathbf{B}_l,\mathbf{Q}_l}(-\alpha_l) - m^{\circ}_{\mathbf{B}_l,\mathbf{Q}_l}(-\alpha_l) \asymp 0 \quad (83)$$

*where $m^{\circ}_{\mathbf{B}_l,\mathbf{Q}_l}(-\alpha_l)$ is given by*

$$m^{\circ}_{\mathbf{B}_l,\mathbf{Q}_l}(-\alpha_l) = \frac{1}{N}\mathrm{tr}\mathbf{Q}_l\mathbf{T}_l \quad (84)$$

*with $\mathbf{T}_l$ is given by*

$$\mathbf{T}_l = \left(\frac{1}{N}\sum_{i=1}^{K}\frac{\mathbf{\Phi}_{lli}}{1 + u_{li}(-\alpha_l)} + \mathbf{S}_l + \alpha_l \mathbf{I}_N\right)^{-1} \quad (85)$$

*where the elements of $\mathbf{u}_l(-\alpha_l) = [u_{l1}(-\alpha_l), \ldots, u_{lK}(-\alpha_l)]^T$ are defined as $u_{li}(-\alpha_l) = \lim_{t \to \infty} u_{li}^{(t)}(-\alpha_l)$, where for $t \in \{1, 2, \ldots\}$*

$$u_{lk}^{(t)}(-\alpha_l) = \frac{1}{N}\mathrm{tr}\mathbf{\Phi}_{llk}\left(\frac{1}{N}\sum_{i=1}^{K}\frac{\mathbf{\Phi}_{lli}}{1 + u_{li}^{(t-1)}(-\alpha_l)} + \mathbf{S}_l + \alpha_l \mathbf{I}_N\right)^{-1} \quad (86)$$

*with initial values $u_{lk}^{(0)}(-\alpha) = \frac{1}{\alpha}$ for all $k$.*

**Theorem 8.** *[16] Let $\mathbf{\Omega}_l \in \mathbb{C}^{N \times N}$ be Hermitian nonnegative definite with uniformly bounded spectral norm (with respect to $N$). Under the conditions of Theorem 1*

$$\frac{1}{N}\mathrm{tr}\mathbf{Q}_l(\frac{1}{N}\hat{\mathbf{H}}_{ll}\hat{\mathbf{H}}_{ll}^H + \mathbf{S}_l + \alpha_l \mathbf{I}_N)^{-1}\mathbf{\Omega}_l \quad (87)$$
$$(\frac{1}{N}\hat{\mathbf{H}}_{ll}\hat{\mathbf{H}}_{ll}^H + \mathbf{S}_l + \alpha_l \mathbf{I}_N)^{-1} - \frac{1}{N}\mathrm{tr}\mathbf{Q}_l\mathbf{T}'_{l,\mathbf{\Omega}_l} \asymp 0$$

*where $\mathbf{T}'_{l,\mathbf{\Omega}_l} \in \mathbb{C}^{N \times N}$ is defined as*

$$\mathbf{T}'_{l,\mathbf{\Omega}_l} = \mathbf{T}_l \times \left(\frac{1}{N}\sum_{j=1}^{K}\frac{u'_{lj,\mathbf{\Omega}_l}(-\alpha_l)\mathbf{\Phi}_{llj}}{(1 + u_{lj}(-\alpha_l))^2} + \mathbf{\Omega}_l\right) \times \mathbf{T}_l \quad (88)$$

*where $\mathbf{T}_l$ and $\mathbf{u}_l(-\alpha)$ are given by theorem 1, and $\mathbf{u}'_{l,\mathbf{\Omega}_l}(-\alpha) = [u'_{l1,\mathbf{\Omega}_l}(-\alpha), \ldots, u'_{lK,\mathbf{\Omega}_l}(-\alpha)]^T$ is computed from*

$$\mathbf{u}'_{l,\mathbf{\Omega}_l}(-\alpha) = (\mathbf{I}_K - \mathbf{J}_l)^{-1}\mathbf{v}_{l,\mathbf{\Omega}_l} \quad (89)$$

*where $\mathbf{J}_l \in \mathbb{C}^{K \times K}$ and $\mathbf{v}_l \in \mathbb{C}^K$ are:*

$$[\mathbf{J}_l]_{mn} = \frac{\mathrm{tr}\mathbf{\Phi}_{llm}\mathbf{T}_l\mathbf{\Phi}_{lln}\mathbf{T}_l}{N^2(1 + u_{ln}(-\alpha))^2} \quad 1 \le m, n \le K \quad (90)$$

$$[\mathbf{v}_{l,\mathbf{\Omega}_1}]_{t1} = \frac{1}{N}\mathrm{tr}\mathbf{\Phi}_{llt}\mathbf{T}_l\mathbf{\Omega}_l\mathbf{T}_l \quad 1 \le t \le K \quad (91)$$

**Lemma 2** (Matrix Inversion Lemma). *Let $\mathbf{U}$ be an $N \times N$ invertible matrix and $x \in \mathbb{C}^N$, $c \in \mathbb{C}$ for which $\mathbf{U} + c\,\mathbf{xx}^H$ is invertible. Then*

$$\mathbf{x}^H\left(\mathbf{U} + c\,\mathbf{xx}^H\right)^{-1} = \frac{\mathbf{x}^H\mathbf{U}^{-1}}{1 + c\,\mathbf{x}^H\mathbf{U}^{-1}\mathbf{x}}. \quad (92)$$

**Lemma 3** (Trace Lemma). *Let $\mathbf{A} \in \mathbb{C}^{N \times N}$ and $\mathbf{x}, \mathbf{y} \sim \mathcal{CN}(\mathbf{0}, \frac{1}{N}\mathbf{I}_N)$. Assume that $\mathbf{A}$ has uniformly bounded spectral norm (with respect to $N$) and that $\mathbf{x}$ and $\mathbf{y}$ are mutually independent and independent of $\mathbf{A}$. Then, for all $p \ge 1$,*

$$\mathbf{x}^H\mathbf{A}\mathbf{x} - \frac{1}{N}\mathrm{tr}\mathbf{A} \asymp 0 \quad and \quad \mathbf{x}^H\mathbf{A}\mathbf{y} \asymp 0. \quad (93)$$

**Lemma 4** (Rank-1 perturbation lemma). *Let $\mathbf{A}_1$, $\mathbf{A}_2$, ..., with $\mathbf{A}_N \in \mathbb{C}^{N \times N}$, be deterministic with uniformly bounded spectral norm and $\mathbf{B}_1$, $\mathbf{B}_2$, ..., with $\mathbf{B}_N \in \mathbb{C}^{N \times N}$, be random Hermitian, with eigenvalues $\lambda_1^{\mathbf{B}_N} \le \lambda_2^{\mathbf{B}_N} \le \ldots \le \lambda_N^{\mathbf{B}_N}$ such that, with probability 1, there exist $\epsilon > 0$ for which $\lambda_1^{\mathbf{B}_N} > \epsilon$ for all large $N$. Then for $\mathbf{v} \in \mathbb{C}^N$*

$$\frac{1}{N}\mathrm{tr}\mathbf{A}_N\mathbf{B}_N^{-1} - \frac{1}{N}\mathrm{tr}\mathbf{A}_N(\mathbf{B}_N^{-1} + \mathbf{vv}^H)^{-1} \asymp 0. \quad (94)$$

*where $\mathbf{B}_N^{-1}$ and $(\mathbf{B}_N^{-1} + \mathbf{vv}^H)^{-1}$ exist with probability 1.*